\pdfoutput=1
\documentclass[runningheads]{llncs}

\usepackage{graphicx}
\usepackage{amsmath, amssymb}
\usepackage{booktabs}
\usepackage{subcaption}
\usepackage{hyperref}
\usepackage{enumitem}
\usepackage{bbding}

\usepackage[section]{placeins}

\usepackage{xurl}
\usepackage{microtype}
\newcommand{\code}[1]{\texttt{\nolinkurl{#1}}}

\begin{document}

\title{SafeWall: A Risk-Aware Job Walltime Recommendation for HPC Scheduling\thanks{Accepted at the 29th Workshop on Job Scheduling Strategies for Parallel Processing (JSSPP 2026), held with IEEE IPDPS 2026, New Orleans, LA, USA. This is the authors' accepted manuscript; the final version will appear in Springer Lecture Notes in Computer Science.}}
\titlerunning{SafeWall: Risk-Aware Walltime Recommendation}

\author{
Kanglin Xu\inst{1}\and
Michael Papka\inst{1,2} \and
Zhiling Lan\inst{1,2}
}

\authorrunning{K. Xu et al.}

\institute{
University of Illinois Chicago, Chicago, IL, USA\\
\texttt{\{kxu27,papka,zlan\}@uic.edu}
\and
Argonne National Laboratory, Lemont, IL, USA
}

\maketitle

\begin{abstract}
In high-performance computing (HPC), users must submit resource requirements for their jobs, with user-supplied walltime being a critical parameter. However, these estimates are notoriously inaccurate, as users tend to overestimate walltime to prevent the scheduler from terminating their jobs prematurely. While existing work on learning-based walltime prediction primarily focuses on maximizing prediction accuracy, such predictions are inherently risky for walltime recommendations. Specifically, underestimation leads to job termination before completion, resulting in lost progress and wasted resources. In this study, we present SafeWall, a novel recommendation method that adjusts walltime predictions by incorporating an explicit, nonnegative safety buffer to mitigate underestimation risk. The core of our approach is a risk-aware buffer that dynamically calibrates safety margins based on individual user's historical behavior. When user data is limited, the system employs statistical shrinkage to align these margins with a global baseline, ensuring robustness. This yields a lightweight mechanism that optimizes the trade-off between prediction accuracy and termination risk. SafeWall is modular and applicable to various learning-based prediction models. We evaluate SafeWall using five public workload traces, including the full-lifetime job log from the 4,392-node Theta machine at the Argonne Leadership Computing Facility. Across all traces, SafeWall substantially reduces underestimation risk compared to the unbuffered base predictors, while preserving competitive accuracy. The source code is publicly available at \url{https://github.com/SPEAR-UIC/JSSPP26-SafeWall}.
\end{abstract}


\section{Introduction}

Batch schedulers in high-performance computing (HPC) systems rely on runtime information for key scheduling decisions, especially backfilling, where the scheduler opportunistically fits shorter jobs into reservation gaps. However, schedulers typically observe only the user-supplied walltime at submission; this walltime is often a conservative upper bound rather than a calibrated estimate. This mismatch creates a dilemma: an overly large walltime reduces backfilling opportunities, while a small walltime may cause premature job termination, wasting all previously performed computation. Prior work has shown that replacing raw walltime with system-generated runtime information can improve scheduler behavior compared to directly using user-supplied walltime in backfilling settings \cite{srinivasan_characterization_2002,mualem_utilization_2001,tsafrir_backfilling_2007}.

Various runtime prediction techniques have been presented for refining walltime recommendations, including both classical statistical methods and machine learning-based methods.
Recent work explores a range of predictors, including ensemble learning \cite{chen2020ensemble} and transformer-style architectures \cite{chen2023pctransformer}, and shows that richer application-level signals can improve runtime prediction when available \cite{lamar2023appparams}. However, \emph{high prediction accuracy does not translate into safe walltime recommendations}. Batch scheduling imposes a strict penalty for underestimation: if a recommendation is smaller than the actual runtime, the job is terminated; in contrast, a similarly sized overestimate typically only increases reserved slack.
Fan et al.\ \cite{fan2017tradeoff} formalize this asymmetry by emphasizing the trade-off between accuracy and underestimation rate (UR), arguing for evaluation that measures underestimation risk rather than only prediction error. This highlights a deployment-facing gap: a state-of-the-art predictor optimized for symmetric loss functions (penalizing under- and over-prediction equally) can yield high prediction accuracy yet fail as walltime recommendations by producing a large fraction of underestimates.

In this work, we treat walltime recommendation as a risk-control problem. Specifically, we introduce \textit{SafeWall}, a two-component design that separates estimation from safety. SafeWall first trains a runtime estimator using scheduler-visible features (\emph{base predictor}); it then applies a model-agnostic risk-aware adapter that produces a nonnegative safety margin (\emph{risk-aware buffer}) and outputs a conservative walltime recommendation. This separation is advantageous because it decouples predictive accuracy from risk management: improvements to the base predictor directly translate to better recommendations without redesigning the safety mechanism, while the adapter can enforce explicit underestimation controls independently of the underlying model. As a result, SafeWall is modular, portable across predictors, and robust to changes in workload characteristics.
Moreover, the separation makes SafeWall complementary to prior efforts on runtime prediction: any predictor that outputs a runtime estimate can be wrapped with the same buffering logic, allowing existing models to be reused without modification.

A key insight that inspired SafeWall is that underestimation risk is heterogeneous across users. Users exhibit distinct submission patterns and habits; hence, a single global safety margin can be unnecessarily conservative for most jobs while still unsafe for specific users. To address this, SafeWall introduces a user-specific safety buffer: it learns each user's degree of conservatism from historical data and uses this information to adjust the risk-aware buffer. In cases of cold start and low activity, SafeWall shrinks the risk-aware buffer toward a global baseline, avoiding unstable or overly aggressive personalization. This yields a continuum between global and personalized risk control, providing stability for cold-start and low-activity users while exploiting user-level structure when sufficient evidence exists.

We first evaluate SafeWall using the full-lifetime
job log from the 4,392-node Theta machine at the Argonne Leadership
Computing Facility \cite{alcf_theta}. In our experiments, we use XGBoost, a gradient-boosted decision-tree regressor for tabular scheduler logs, as the base predictor due to its strong empirical performance in prior runtime prediction studies \cite{chen2016xgboost,menear2023mastering}.
We observe that high predictive accuracy does not necessarily translate to a ``safe'' recommendation; even with a strong regression fit, the base predictor may suffer from a high underestimation rate. By incorporating an explicit nonnegative buffer, SafeWall significantly reduces underestimation risk. As expected, more conservative buffers reduce the underestimation rate at the cost of increased prediction slack.

We further evaluate the robustness of SafeWall using multiple public workload traces collected from established repositories, such as the Parallel Workloads Archive \cite{pwa_site,klusacek2024trace,klusacek2023softwalltime,nerscCoriDocs}. Our results demonstrate that the risk-control layer introduced by SafeWall consistently reduces underestimation risk while maintaining prediction accuracy comparable to existing learning-based methods.

Overall, this work offers the following key contributions:
\begin{itemize}
    \item We introduce SafeWall, a practical and transparent walltime recommendation design that separates runtime estimation from risk control via an explicit nonnegative buffering layer that can wrap any base predictor (\S\ref{sec:methodology}).

    \item We evaluate SafeWall using production scheduler logs from multiple systems to assess its robustness across diverse architectures and workloads (\S\ref{sec:results}).

    \item We discuss the strengths and limitations of SafeWall (\S\ref{sec:discussion}).
\end{itemize}

The source code is publicly available at \url{https://github.com/SPEAR-UIC/JSSPP26-SafeWall}.

\section{Related Work}
\label{sec:related}

Runtime estimates are a core scheduling primitive in HPC batch systems, informing decisions such as prioritization, queue management, and backfilling. In practice, however, schedulers typically rely on user-provided walltime limits, which are often discretized and conservative upper bounds rather than accurate runtime predictions. Using these values as if they were true processing times can degrade system performance; yet, replacing them with learned predictions raises an important operational concern: \emph{underestimation} can be catastrophic because jobs may be terminated at the walltime limit. Fan et al.~\cite{fan2017tradeoff} emphasize that evaluation must therefore go beyond average error and explicitly quantify one-sided risk (e.g., the underestimation rate), since improving standard accuracy measures can inadvertently increase termination risk. Related work on underestimation-aware analysis includes an erratum notice associated with a prior study line on underestimation detection~\cite{guo2018erratum}, underscoring that measurement and evaluation details matter.

A large body of work focuses on improving the base predictor $\hat{y}$ by exploring different model families and feature sets. Ensemble learning has been shown to improve predictive accuracy over single models in HPC runtime prediction~\cite{chen2020ensemble}, and transformer-style architectures have been proposed to capture complex nonlinear dependencies in scheduler logs~\cite{chen2023pctransformer}. Another recurring theme is feature availability: although richer application-level signals can improve prediction, they are often difficult to collect and standardize across workloads and systems. Lamar et al.~\cite{lamar2023appparams} quantify the value of application input parameters, highlighting the potential gains from domain-specific features but also the practical barriers to deploying such approaches broadly in production settings. These results motivate methods that achieve strong performance using scheduler-visible features and lightweight historical summaries.

Beyond base prediction, some approaches integrate runtime inference into the scheduling loop. Zrigui et al.~\cite{zrigui2022onlineclass} propose online job runtime classification to continuously refine processing-time information used by the scheduler, demonstrating that classification-oriented runtime information can improve scheduling outcomes compared to naively using loose walltime requests. In parallel, several systems incorporate conservative margins or post-processing rules into walltime recommendations to mitigate harmful underprediction. Oztop et al.~\cite{oztop_jobgrouping} propose a job grouping framework that clusters similar jobs before prediction and discuss conservative strategies to reduce underestimation risk, reinforcing the view that post-processing policies can remain valuable even when base predictors are strong. Similarly, Cui et al.~\cite{cui2025clustering} target backfilling with a classification-oriented pipeline and conservative recommendations to reduce operational underprediction.

Finally, the realism of evaluation depends on trace fidelity and label quality: ``runtime'' recorded in logs may include artifacts (e.g., measurement granularity, truncation, or missing/censored outcomes). Klus{\'a}{\v{c}}ek and Chlumsk{\'y}~\cite{klusacek2024trace} introduce a real-life workload trace with refined runtime estimates and show that improving labels can materially change conclusions about prediction and scheduling policies. This highlights an important practical consideration for many studies: even strong models can be shaped by the quality and semantics of the available runtime labels.

\emph{The novelty of SafeWall} lies in its separation of prediction from recommendation. While prior work focuses primarily on maximizing accuracy through advanced models or richer feature sets, SafeWall introduces an explicit risk-control layer atop any base predictor. By decoupling runtime estimation from recommendation via a model-agnostic, nonnegative buffer that adapts to individual historical behaviors, SafeWall remains orthogonal to existing research. It serves as a complementary framework that enhances, rather than replaces, traditional walltime prediction studies.

\section{Methodology}
\label{sec:methodology}

HPC clusters typically use batch schedulers in which users submit jobs and the scheduler decides when and where they run; each job is submitted with a walltime limit specifying its maximum allowed runtime.
The objective of SafeWall is twofold: (i) to accurately estimate a job's actual execution time, and (ii) to recommend a conservatively buffered walltime that minimizes the risk of premature termination.

\begin{figure}[t]
\centering
\includegraphics[width=\textwidth]{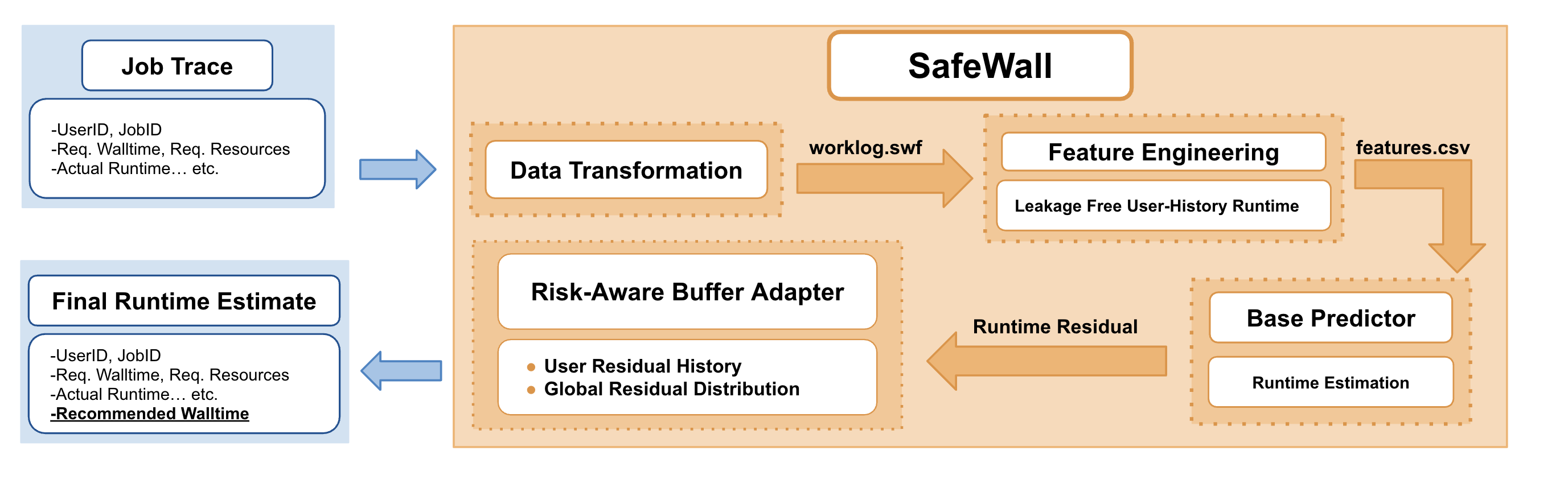}
\caption{SafeWall overview. The pipeline consists of a base predictor that estimates job runtime and a risk-aware buffer adapter that adjusts runtime estimation based on user-specific historic information.}
\label{fig:workflow}
\end{figure}

As shown in Figure~\ref{fig:workflow}, SafeWall adopts a two-component design. First, a learning-based runtime estimator produces a runtime estimate from scheduler-visible features. Second, a model-agnostic risk-aware adapter computes a nonnegative adjustment value (i.e., buffer) from historical prediction outcomes and outputs the recommended walltime. SafeWall first trains a base predictor on scheduler-visible features (and lightweight user-history summaries when available) to produce $\hat{y}$. It then calibrates a buffer from training residuals $e=y-\hat{y}$ and applies a nonnegative adjustment $b$---globally or in a risk-aware, shrinkage-stabilized manner---to form the final recommendation $\tilde{y}=\hat{y}+b$.

Because this buffering step is applied after prediction and is calibrated using past prediction errors, it is compatible with the various runtime predictors described in Section~\ref{sec:related} (e.g., ensembles or transformer-based models). This makes SafeWall complementary to existing efforts that focus solely on improving the base predictor accuracy.

Formally, let $i \in \{1,\dots,n\}$ index the jobs. Each job is associated with a feature vector $\mathbf{x}_i \in \mathbb{R}^d$ available at submission time (where $d$ is the number of features), an observed runtime $y_i \in \mathbb{R}_{\ge 0}$ (seconds), and a user identifier $u_i$ used for optional risk-aware buffering. SafeWall learns a prediction function $f(\cdot)$ that produces a runtime estimate
\begin{equation}
\hat{y}_i = f(\mathbf{x}_i),
\end{equation}
and then applies a nonnegative safety buffer to convert this estimate into a conservative walltime recommendation:
\begin{equation}
\tilde{y}_i = \hat{y}_i + b_i, \qquad b_i \ge 0.
\end{equation}

\subsection{Data Transformation}
\label{subsec:preprocess}

HPC schedulers typically archive workload trace data in disparate formats and with inconsistent field definitions. A rigorous preprocessing pipeline is therefore essential to transform raw logs into a unified structure suitable for training walltime recommendation models.

In this study, we utilize a production workload trace from the Theta supercomputer at the Argonne Leadership Computing Facility (ALCF) \cite{alcf_theta}. We perform several preprocessing steps to transform the raw log into a supervised learning dataset, which includes standardizing fields and adjusting right-censored records (i.e., jobs that did not complete normally).

\begin{table}[t]
\centering
\caption{Mapping from Theta log fields to Standard Workload Format (SWF) fields. SWF fields not available in the log are set to $-1$.}
\label{tab:swf-mapping}
\small
\setlength{\tabcolsep}{6pt}
\renewcommand{\arraystretch}{1.1}
\begin{tabular}{ll}
\toprule
\textbf{SWF Field} & \textbf{Theta Log Field} \\
\midrule
Submit time & queued timestamp normalized to seconds since trace start \\
Wait time & \code{ELIGIBLE_WAIT_SECONDS} \\
Run time & \code{RUNTIME_SECONDS} \\
Requested time (walltime) & \code{WALLTIME_SECONDS} \\
Allocated nodes & \code{NODES_USED} \\
Exit status & \code{EXIT_STATUS} \\
User identifier & \code{user} \\
Project identifier & \code{group\_id} \\
\bottomrule
\end{tabular}
\end{table}

\begin{itemize}
    \item \textbf{Workload log $\rightarrow$ unified trace format.}
    We map core workload log fields (submit time, wait time, runtime, requested walltime, nodes, exit status, and user/project identifiers) into Standard Workload Format (SWF) to ensure consistent definitions and reproducibility \cite{swfbenchmarks}. Submit times are normalized to seconds since the first job in the trace, and jobs are ordered chronologically (Table~\ref{tab:swf-mapping}).

    \item \textbf{Handle right-censored jobs via runtime capping.}
    Jobs whose recorded runtime reaches or exceeds the requested walltime are effectively right-censored: the true runtime may be longer, but the log provides only the walltime limit. Because SafeWall calibrates buffers from prediction residuals, treating these entries as fully observed would inject unknown truncation into the error distribution. Following common trace-processing practice, we therefore cap the runtime label at the requested walltime for such jobs. This preserves a consistent, scheduler-visible definition of ``observed runtime'' and avoids mixing unbounded overruns with fully observed completion times during model training and buffer estimation.
\end{itemize}

\subsection{Feature Engineering}
\label{subsec:features}

Next, prior to applying any prediction method, we must develop a robust feature representation to capture the essential temporal and structural characteristics of the workload.

We compute per-user lag features from prior completed jobs (previous runtime, second previous runtime, and their mean), along with summary statistics capturing request accuracy and runtime scale. For each prior job $j$, we define request accuracy as
\[
a_j=\min(\text{wall}_j/y_j,\, y_j/\text{wall}_j)\in(0,1],
\]
and include per-user summaries (mean/max) and recent-history variants over the last 10 completed jobs when available. We also include longer-run runtime summaries (mean/max, recent mean/max, and the 25th percentile of prior successful runtime) \cite{fan2017tradeoff,zhang2012kraken_behavior}. All history features are computed strictly from jobs submitted earlier than the current job, preventing information leakage.

\paragraph{Missing values and cold-start users.}
User-history features are undefined for users with insufficient prior completed jobs. Rather than dropping these jobs or imputing missing history, we treat them as cold-start cases and fall back to the user-provided walltime request as the recommendation. This conservative fallback avoids unsafe recommendations when there is not enough evidence to support history-based personalization. After filtering, the dataset contains 430{,}223 jobs; 367{,}477 jobs have sufficient history for the full feature set and are used for model training and evaluation, while 62{,}746 jobs use the walltime fallback. Because batch schedulers enforce a submission-time walltime limit and may terminate jobs when that limit is reached, user-submitted walltime provides a safe default when history features are unavailable \cite{slurm_sbatch}.

\subsection{Base Predictors}
\label{subsec:xgb}

In this study, we consider two gradient-boosted tree models as base predictors:
(i) standard XGBoost trained with squared-error loss for point prediction, and
(ii) a quantile gradient boosting (Quantile-GBR) configuration that directly estimates upper conditional quantiles of runtime.
Tree boosting is a strong baseline for heterogeneous tabular data and effectively captures nonlinear effects and feature interactions
\cite{chen2016xgboost,grinsztajn2022tabular,friedman2001gbm}.

It is important to note that SafeWall is model-agnostic; while we utilize boosted tree models in this study, the buffering mechanism can be seamlessly integrated with other learning-based predictors.

\paragraph{XGBoost (Squared Loss).}
The standard predictor is an additive ensemble of regression trees,
\begin{equation}
\hat{y}_i = \sum_{k=1}^{K} f_k(\mathbf{x}_i),
\qquad f_k \in \mathcal{T},
\end{equation}
where $\mathcal{T}$ denotes CART-style regression trees.
The model is trained to minimize squared error with regularization, as implemented in XGBoost.
Because runtime is nonnegative, we enforce feasibility by clipping predictions:
\begin{equation}
\hat{y}_i \leftarrow \max(\hat{y}_i, 0).
\end{equation}

\paragraph{Quantile Gradient Boosting.}
To directly model conservative runtime estimates, we also train a quantile boosting model that minimizes the pinball loss at quantile level $\tau$:
\begin{equation}
\mathcal{L}_{\tau}(y_i, \hat{y}_i)
=
\begin{cases}
\tau (y_i - \hat{y}_i), & y_i \ge \hat{y}_i, \\
(1-\tau)(\hat{y}_i - y_i), & y_i < \hat{y}_i.
\end{cases}
\end{equation}
For $\tau > 0.5$, the model emphasizes underprediction penalties and produces upper-quantile runtime estimates.
This configuration provides a direct alternative to post-hoc buffering.

For both models, we tune hyperparameters using randomized search with five-fold cross-validation on the training split.
We select the configuration that maximizes mean cross-validated $R^2$ (for squared-loss models) or minimizes quantile loss (for quantile models), and then retrain the final model on the full training set.
The resulting predictions on the training and test splits are subsequently passed to the SafeWall buffering policies described in Section~\ref{subsec:buffer}.

\subsection{Risk-Aware Buffer Adapter}
\label{subsec:buffer}

A base prediction $\hat{y}_i$ may achieve excellent accuracy while still being operationally risky for batch scheduling due to asymmetric costs: an \emph{underestimate} can terminate a job at the walltime limit, whereas an \emph{overestimate} typically results only in reserved slack.

\begin{figure}
    \centering
    \includegraphics[width=0.5\textwidth]{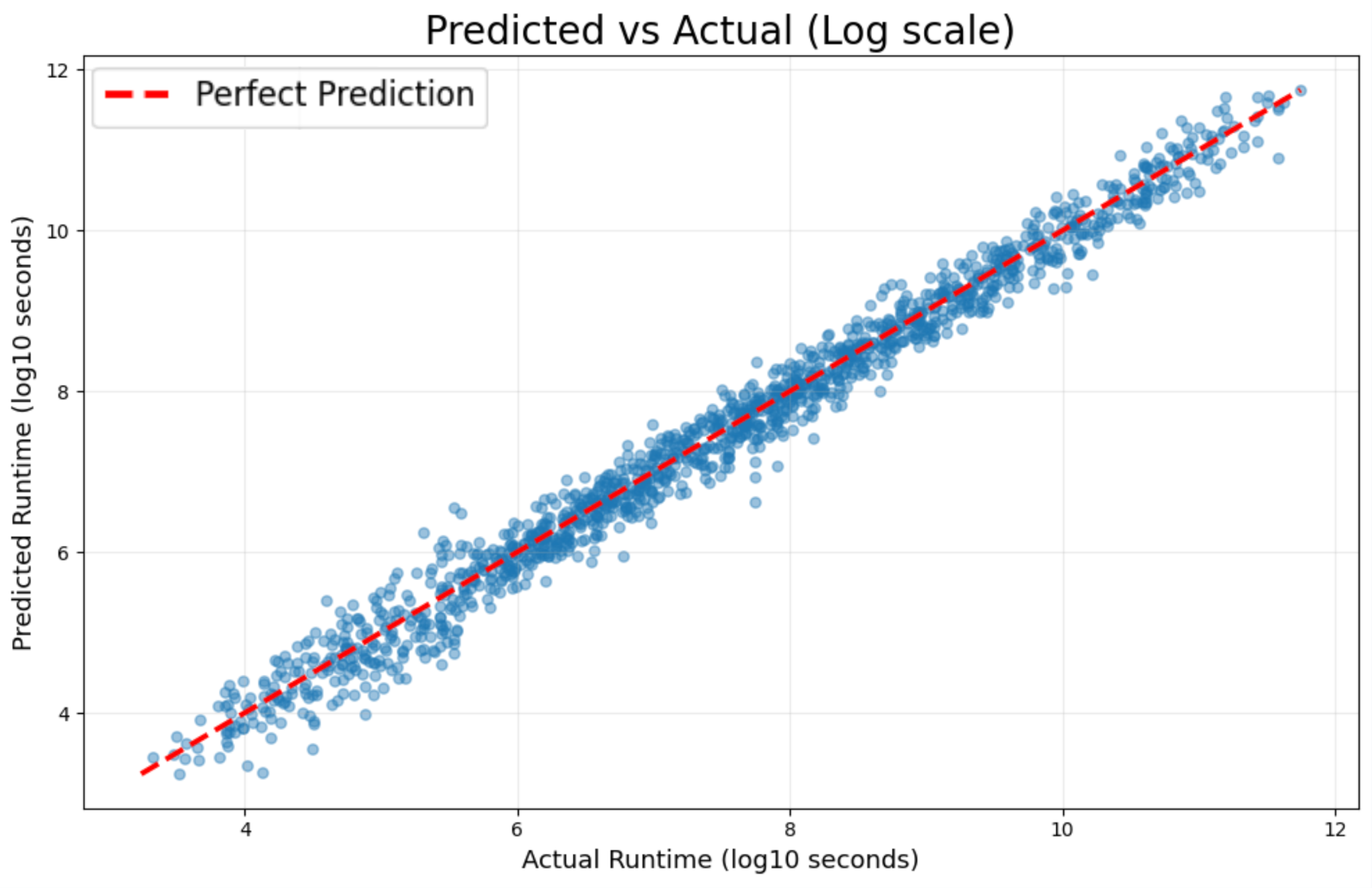}
    \caption{Illustrative comparison of predicted versus actual job runtimes on a log scale.}
    \label{fig:ur_regression_illustration}
\end{figure}

Figure~\ref{fig:ur_regression_illustration} presents an illustrative example. Because the predictor (XGBoost) is trained with a symmetric loss, residuals are approximately balanced around the ideal line $y=\hat{y}$, so points are distributed on both sides of the diagonal. Jobs lying above the diagonal ($y>\hat{y}$) are underestimated and therefore at risk of termination; when errors are roughly symmetric, this can yield underestimation rates approaching 50\% even when global fit is strong. Accordingly, improving base prediction accuracy alone does not guarantee scheduler safety. We therefore treat walltime recommendation as a risk-control layer applied atop the base prediction, introducing an explicit upward adjustment to systematically reduce the mass of underestimates.

Let $y_i$ denote the true runtime (seconds) and $\hat{y}_i=f(\mathbf{x}_i)$ the base estimate from the trained regressor. Define the residual
\begin{equation}
e_i = y_i - \hat{y}_i,
\end{equation}
so $e_i>0$ indicates underprediction. The buffer generates a nonnegative adjustment value $b_i\ge 0$ and produces a conservative walltime recommendation
\begin{equation}
\tilde{y}_i = \hat{y}_i + b_i.
\label{eq:buffered_pred}
\end{equation}
To evaluate and tune buffering policies, we focus on the scheduler-facing trade-off between termination risk and conservatism.

Termination risk is captured by the underestimation rate (UR). After buffering, define the post-buffer residual as $r_i = y_i-\tilde{y}_i$. A job is underestimated when $r_i>0$, meaning the recommended walltime is shorter than the actual runtime and the job would be at risk of hitting the walltime limit under hard termination. We therefore compute UR using Eq.~\eqref{eq:UR} as the fraction of jobs with $r_i>0$.

SafeWall derives an additive safety margin from residuals and optionally personalizes it when there is sufficient user history. The design begins with a globally calibrated error buffer and then applies a user-adjusted tweak when historical evidence supports personalization. This keeps the mechanism simple and auditable while explicitly targeting the one-sided underprediction tail that drives termination risk. Specifically, SafeWall proceeds in three steps as detailed below.

\noindent\emph{Step 1: global error buffer from training residuals.}
We begin by calibrating a single global adjustment value using the model's historical underprediction errors. After training the point predictor, we run it on the training set and record the residual for each job,
\[
e_i = y_i-\hat{y}_i,
\]
which measures how far the prediction is below (or above) the realized runtime. Positive residuals ($e_i>0$) are precisely the cases that matter for termination risk: they correspond to jobs whose true runtime exceeded the point estimate.

To translate these historical underprediction errors into a conservative shift, we choose a target underestimation rate $\tau\in(0,1)$ (e.g., $\tau=0.10$). Intuitively, $\tau$ specifies the fraction of future jobs we are willing to allow to still exceed the recommended walltime after buffering. We then set the global adjustment value $b_{\mathrm{g}}(\tau)$ to a high quantile of the positive residual distribution:
\begin{equation}
b_{\mathrm{g}}(\tau)
=
Q_{1-\tau}\!\left(\{\, e_i : e_i>0,\ i\in\mathcal{I}_{\text{train}} \,\}\right).
\label{eq:global_pos_buffer}
\end{equation}
This choice has a direct interpretation. Consider all training jobs for which the model underpredicted. By construction, $b_{\mathrm{g}}(\tau)$ is large enough to cover approximately a $(1-\tau)$ fraction of those historical underpredictions; that is, for most past underpredicted jobs, $\hat{y}_i + b_{\mathrm{g}}(\tau)$ would have met or exceeded $y_i$. When training and test conditions are similar, applying the same adjustment on new jobs aims to keep the empirical underestimation rate near $\tau$, while exposing a clear knob that trades lower UR for higher reserved slack.

\noindent\emph{Step 2: user-adjusted tweak with shrinkage.}
A single global adjustment value is an average-case policy. In practice, residual behavior can be heterogeneous across users: some users routinely run longer than the model predicts (large positive residuals), while others are typically well covered by the point predictor or even overpredicted. To reflect this heterogeneity, we estimate a risk-aware adjustment value from each user's historical underprediction residuals, and then stabilize it using shrinkage.

For each user $u$, let $\mathcal{E}_u=\{\,e_i : e_i>0,\ u_i=u,\ i\in\mathcal{I}_{\text{train}}\,\}$ denote the set of that user's positive training residuals, and let $n_u = |\mathcal{E}_u|$ be the number of usable underprediction cases. Using the same target underestimation rate $\tau$, we define a risk-aware residual quantile
\[
b_u(\tau)=Q_{1-\tau}(\mathcal{E}_u),
\]
so users with larger or more frequent underpredictions naturally receive a larger adjustment.

However, $b_u(\tau)$ can be noisy when $n_u$ is small (e.g., new or infrequent users). To avoid unstable jumps caused by limited evidence, we shrink the risk-aware estimate toward the global baseline $b_{\mathrm{g}}(\tau)$ using a convex combination:
\begin{equation}
\tilde{b}_u(\tau)
=
\frac{n_u}{n_u+\kappa}\, b_u(\tau)
+
\frac{\kappa}{n_u+\kappa}\, b_{\mathrm{g}}(\tau),
\label{eq:shrinkage}
\end{equation}
where $\kappa>0$ is a smoothing parameter that determines how much history is needed before personalization dominates. This weighting has a simple interpretation: when $n_u$ is large, the first term approaches $b_u(\tau)$ and the adjustment becomes strongly risk-aware; when $n_u$ is small, the second term dominates and the method falls back toward the global adjustment. In the extreme cold-start case $n_u=0$, the rule reduces exactly to $\tilde{b}_u(\tau)=b_{\mathrm{g}}(\tau)$.

\noindent\emph{Step 3: apply the adjustment value.}
For a job from user $u_i$, the recommended walltime is
\begin{equation}
\tilde{y}_i = \hat{y}_i + \max\{0,\tilde{b}_{u_i}(\tau)\}.
\label{eq:user_buffered_pred}
\end{equation}

\noindent\emph{Rolling window option.}
To accommodate drift, we also consider computing $b_u(\tau)$ from only the most recent $W$ completed jobs, while keeping the same shrinkage rule toward $b_{\mathrm{g}}(\tau)$.

\section{Evaluation}
\label{sec:results}

\subsection{Experimental Configuration}

\begin{figure}[t]
    \centering
    \includegraphics[width=0.85\textwidth]{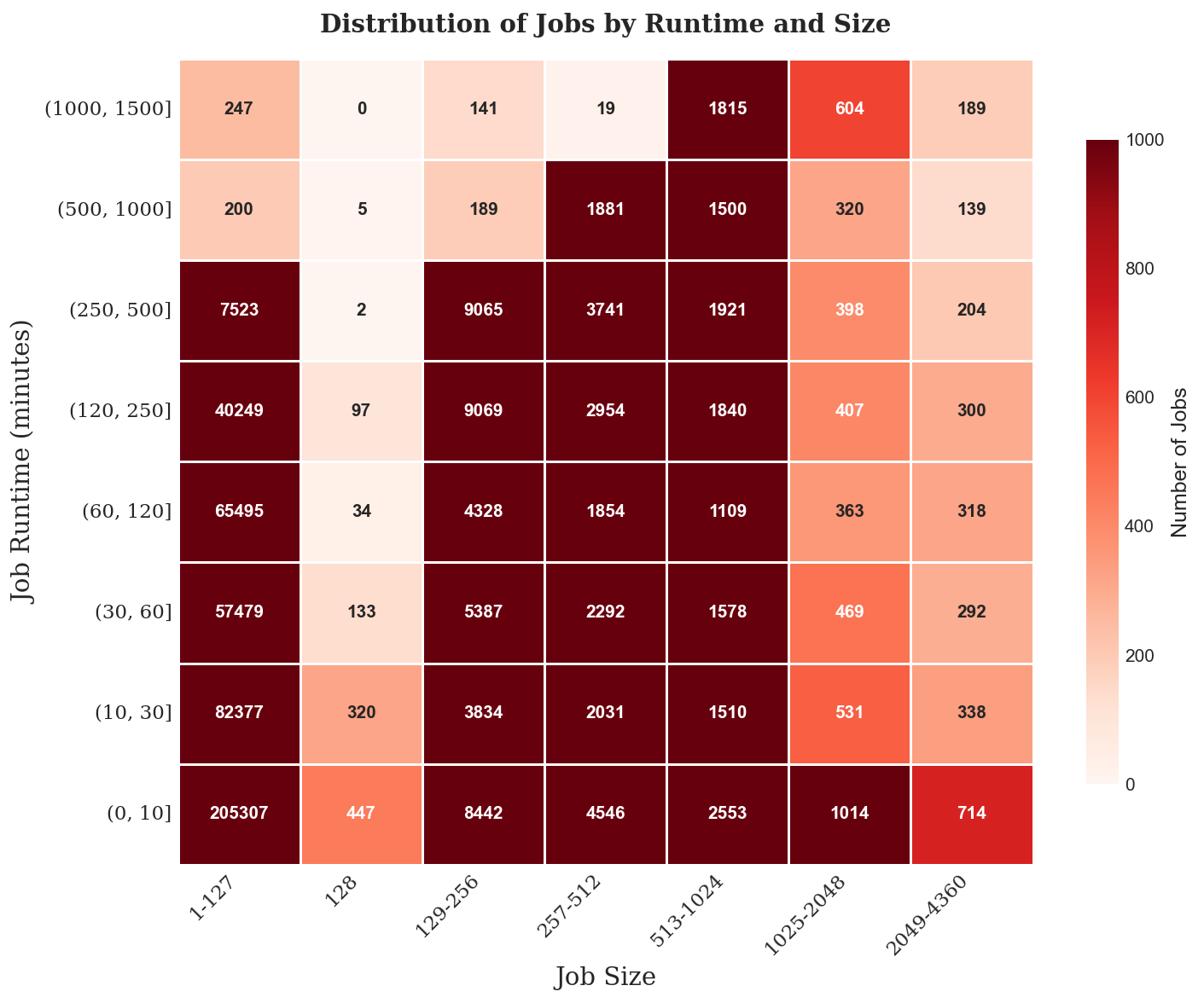}
    \caption{Joint distribution of job runtime and resource scale for the 6.5-year Theta workload. Each cell represents the frequency of completed jobs within specific runtime (rows) and node-count (columns) bins; color intensity is capped at 1{,}000 jobs for visual clarity.}
    \label{fig:theta-runtime-size}
\end{figure}

\textbf{Workload Traces.}
We evaluate SafeWall using production scheduler logs from multiple systems to assess its robustness across diverse architectures and workloads. These include the Theta system at ALCF (\S\ref{subsec:results_overall}), as well as several supercomputers from publicly available traces (\S\ref{subsec:robustness}).
In the remainder of the paper, we use the first 80\% of each workload trace for model training and report results based on the remaining 20\%.

The Theta log spans approximately 6.5 years, from June~2017 to December~2023, and comprises 540{,}503 batch jobs submitted by 1{,}827 unique users across 628 projects, accounting for a total of 13.42 billion core hours \cite{alcf_theta}. For evaluation, we partition the trace into an 80\% training split and a 20\% held-out test split, ensuring that the predictor and buffer are calibrated using only past jobs relative to the test period.

As shown in Figure~\ref{fig:theta-runtime-size}, the Theta workload is strongly skewed toward small-scale, short-duration jobs: roughly two-thirds of all submissions use 1--127 nodes (about 65\%), and a majority finish within 30 minutes (about 58\%, with a median runtime of $\sim$18 minutes). Multiple bins in this low-node/short-runtime region exceed the heatmap's capped color scale (1{,}000 jobs per bin), consistent with a workload mix dominated by debugging, testing, and other short iterative runs. The allocation portfolio is largely Discretionary (about 72\%), and users submit conservative walltime requests (median utilization/runtime-to-walltime ratio of about 36.5\%), yet the scheduler still keeps the dominant small-job regime responsive (median queue wait of $\sim$1.4 minutes). In contrast, mid-to-large scale jobs (129--1{,}024 nodes) exhibit a clearer diagonal density pattern with concentrations in multi-hour runtime bands (roughly 120--500 minutes), characteristic of production-grade simulations that scale both runtime and resource use.

A notable feature is a pronounced mass around 128 nodes (about 20\% of jobs), indicating a commonly chosen request size rather than a sparse boundary. Finally, the largest capability-scale jobs (2{,}049--4{,}360 nodes) are rare (on the order of $\sim$0.5\% of submissions) but span long runtimes (up to $\sim$31 hours) and account for a disproportionate share of total core-hours (on the order of $\sim$15\%), forming the resource-intensive tail of the workload.

\noindent\textbf{Evaluation Metrics.}
Our evaluation focuses on the primary objective of SafeWall: (i) lowering termination risk from underestimated walltimes and (ii) maintaining tight recommendations (i.e., avoiding unnecessary over-reservation).

Let $y_i$ denote the true runtime and $\tilde{y}_i$ the recommended walltime produced by a method (for the no-buffer baseline, $\tilde{y}_i=\hat{y}_i$).
We define the post-recommendation residual $r_i = y_i - \tilde{y}_i$. When $r_i>0$, the recommendation is an underestimate and would be at risk of hitting a hard walltime limit. We quantify this risk using the underestimation rate (UR) in Eq.~\eqref{eq:UR}. To quantify how \emph{tight} the recommendation is around the true runtime (regardless of sign), we use the symmetric ratio accuracy in Eq.~\eqref{eq:ratio_accuracy}. Finally, to capture the magnitude of these errors, we also report standard regression metrics (RMSE, MAE, and $R^2$) in Eqs.~\eqref{eq:rmse}--\eqref{eq:r2}; unless otherwise noted, these are computed between $\tilde{y}_i$ and $y_i$ to reflect the quality of the final recommendation.

\begin{equation}
\mathrm{UR} =
\frac{\text{Number of Underpredicted Jobs}}{\text{Number of Total Jobs}}
\times 100\%.
\label{eq:UR}
\end{equation}

\begin{equation}
\mathrm{Accuracy} \;=\; \frac{\min\{\tilde{y}_i,\, y_i\}}{\max\{\tilde{y}_i,\, y_i\}}.
\label{eq:ratio_accuracy}
\end{equation}

\begin{equation}
\mathrm{RMSE} \;=\; \sqrt{\frac{1}{N}\sum_{i=1}^{N}\big(\tilde{y}_i - y_i\big)^2}.
\label{eq:rmse}
\end{equation}

\begin{equation}
\mathrm{MAE} \;=\; \frac{1}{N}\sum_{i=1}^{N}\left|\tilde{y}_i - y_i\right|.
\label{eq:mae}
\end{equation}

\begin{equation}
R^2 \;=\; 1 - \frac{\sum_{i=1}^{N}\big(\tilde{y}_i - y_i\big)^2}{\sum_{i=1}^{N}\big(y_i - \bar{y}\big)^2},
\qquad \bar{y}=\frac{1}{N}\sum_{i=1}^{N} y_i .
\label{eq:r2}
\end{equation}

\textbf{Comparison Methods.}
We evaluate SafeWall as a post-processing layer for each base predictor. In this study, we choose two base predictors: XGBoost and Quantile Gradient Boosting Regressor (Quantile GBR) \cite{chen2016xgboost}, where $\tilde{y}_i=\hat{y}_i$ (no post-processing). For SafeWall, it adjusts predicted output by applying a risk-aware, user-specific buffer given by $\tilde{y}_i=\hat{y}_i+b_i$. We compare SafeWall with the corresponding methods:

\begin{itemize}

  \item \emph{Baselines.} We consider two base predictors:
  \begin{enumerate}[label=(\alph*)]
    \item \emph{XGBoost}, a gradient-boosted decision tree regressor trained with squared-error loss to produce point runtime predictions from submission-time features.
    \item \emph{Quantile GBR}, a Gradient Boosting Regression model trained with quantile loss at $q=0.95$, predicts a conservative runtime at the 95th percentile conditional on features (aiming for $\approx$5\% underestimation under good calibration).
  \end{enumerate}

  \item \emph{Naive prediction methods.} We consider three naive buffering methods:
  \begin{enumerate}[label=(\alph*)]
    \item \emph{Percentage buffer}, which inflates each prediction by a fixed factor,
    $\tilde{y}_i = (1+0.2)\hat{y}_i$.

    \item \emph{Fixed-value buffer}, which adds a constant margin of 30 seconds to every prediction,
    $\tilde{y}_i = \hat{y}_i + 30\text{s}$.

    \item \emph{Piecewise adaptive buffer}, which applies a piecewise percentage rule:
    \begin{equation}
    \tilde{y}_i =
    \begin{cases}
    (1+0.10)\hat{y}_i, & \hat{y}_i \le T_1,\\
    (1+0.15)\hat{y}_i, & \hat{y}_i > T_1,
    \end{cases}
    \label{eq:piecewise_buffer}
    \end{equation}
    where the threshold $T_1$ is fixed in advance.
  \end{enumerate}

\end{itemize}

For the risk-aware buffer, we use target underestimation rate $\tau=0.10$ in Eq.~\eqref{eq:global_pos_buffer} and the shrinkage parameter $\kappa=10$ in Eq.~\eqref{eq:shrinkage}. This configuration provides a stable transition from global to personalized buffering while limiting sensitivity to older user behavior. This combination allows us to quantify the benefit of SafeWall's risk-aware, user-adaptive buffering relative to both straightforward heuristics and an established conservative pipeline.

Our evaluation is organized around three questions.
First, \S\ref{subsec:results_overall} quantifies the performance of SafeWall on the Theta log across prediction strategies.
Second, \S\ref{subsec:robustness} evaluates SafeWall on additional workload traces to test SafeWall's robustness across different systems and workloads.
Finally, \S\ref{subsec:user_bias} examines the variance in prediction errors across individual users, demonstrating why personalized calibration improves the risk--tightness trade-off.

\subsection{Results on Theta}
\label{subsec:results_overall}

\begin{table}[tb]
\centering
\caption{Comprehensive comparison of buffering strategies on the Theta log.}
\label{tab:buffer-results-full}
\scriptsize
\setlength{\tabcolsep}{3.5pt}
\renewcommand{\arraystretch}{1.05}
\resizebox{\textwidth}{!}{
\begin{tabular}{l r r r r r}
\toprule
Method & UR (\%) & Accuracy (\%) & RMSE(s) & MAE(s) & $R^2$\\
\midrule\midrule
XGBoost & 47.8489 & 97.20 & 961.44 & 57.34 & 0.9835\\
XGBoost + SafeWall & 4.3051  & 90.27 & 963.89 & 91.81 & 0.9834\\
\midrule
Quantile GBR & 5.5568  & 85.44 & 685.90 & 159.20 & 0.9916\\
Quantile GBR + SafeWall & 3.4614  & 87.70 & 770.25 & 147.55 & 0.9895\\
\midrule\midrule
Percentage buffer & 1.0613 & 82.83 & 1808.66 & 547.43 & 0.9418\\
\midrule
Fixed-value buffer & 7.0521 & 89.91 & 961.85 & 77.54 & 0.9835\\
\midrule
Piecewise buffer & 2.5893 & 82.49 & 1484.90 & 384.39 & 0.9608\\
\bottomrule
\end{tabular}
}
\end{table}

Table~\ref{tab:buffer-results-full} highlights the operational gap between predictive fit and safe walltime recommendations. Despite excellent test fit, using the base predictor directly is unsafe (UR $\approx 47.85\%$). We also include the job-grouping based resource prediction framework of Oztop et al.~\cite{oztop_jobgrouping} by running their released code on the same Theta split. Under the ``resampled\_rf'' configuration with a two-sigma bias setting, the reproduced baseline achieves UR $=3.06\%$ but with much larger error (RMSE $=18{,}505.64$s, MAE $=14{,}733.11$s). In contrast, SafeWall attains a comparable UR in the 3--4\% range (e.g., $4.31\%$ for XGBoost+SafeWall and $3.46\%$ for Quantile+SafeWall) while maintaining substantially lower error (RMSE $\approx 960$--$970$s, MAE $\approx 90$--$150$s), indicating that SafeWall controls termination risk at a similar service level while preserving base-estimation fidelity.

We also compare against a quantile regression baseline implemented as a quantile Gradient Boosting Regressor, which directly predicts an upper conditional runtime quantile as the recommendation. As shown in Table~\ref{tab:buffer-results-full}, SafeWall achieves \emph{lower UR at comparable or better accuracy} than the quantile models at similar operating points, while keeping the underlying point predictor unchanged. In addition, SafeWall is significantly lighter-weight operationally: changing the service level only requires recomputing a residual quantile from the already-trained model, whereas the quantile approach typically requires training (and validating) a separate model for each target quantile.

\emph{Overall, SafeWall should be viewed as a risk-control layer applied on top of an existing runtime predictor, rather than a replacement for the predictor itself, decoupling safety guarantees from predictive modeling.}

\begin{table}[t]
\centering
\caption{Workload traces used for cross-system evaluation. ``Cens.'' denotes the fraction of \emph{censored} jobs whose true runtimes are not observed in the trace (e.g., terminated at the walltime limit or otherwise missing a completion time).}
\label{tab:other_traces}
\scriptsize
\setlength{\tabcolsep}{4pt}
\renewcommand{\arraystretch}{1.10}
\resizebox{\textwidth}{!}{
\begin{tabular}{lrrrrc}
\toprule
Trace & Period & Jobs & Users & Nodes & Cens.~(\%) \\
\midrule
Theta        & Jun 2017--Dec 2023    & 540{,}503      & 1{,}827 & 4{,}392       & 0.1   \\
Cori         & Jan--Dec 2022         & 2{,}462{,}279  & 1{,}318 & 10{,}000      & 0.0  \\
KIT-FH2      & Jun 2016--Jan 2018    & 114{,}024      & 164     & 1{,}173        & 6.6  \\
MetaCentrum  & Jan--Dec 2023         & 10{,}139{,}198 & 1{,}618 & 47 clust.      & 4.9  \\
CERIT-SC-K8s & Jul 2021--Jun 2022    & 64{,}093       & 103     & 20             & 0.0  \\
\bottomrule
\end{tabular}
}
\end{table}

\begin{table}[!htp]
\centering
\caption{Runtime prediction across traces: base model and SafeWall.}
\label{tab:robustness_combined}
\scriptsize
\setlength{\tabcolsep}{4pt}
\renewcommand{\arraystretch}{1.10}
\resizebox{\textwidth}{!}{
\begin{tabular}{l r r r r r}
\toprule
& & \multicolumn{2}{c}{XGBoost}
& \multicolumn{2}{c}{SafeWall} \\
\cmidrule(lr){3-4} \cmidrule(lr){5-6}
Trace
& \#Jobs
& UR (\%)
& Acc (\%)
& UR (\%)
& Acc (\%) \\
\midrule
Cori
& 2{,}465{,}286
& 47.92
& 93.47
& 5.06
& 83.21 \\
KIT
& 114{,}355
& 49.23
& 90.13
& 6.79
& 74.76 \\
CERIT
& 64{,}268
& 49.11
& 92.14
& 5.00
& 83.02 \\
MetaCentrum
& 10{,}145{,}231
& 48.16
& 84.03
& 4.92
& 68.64 \\
\bottomrule
\end{tabular}
}
\end{table}

\begin{figure}
    \centering
    \begin{subfigure}[t]{0.49\textwidth}
        \centering
        \includegraphics[width=\linewidth]{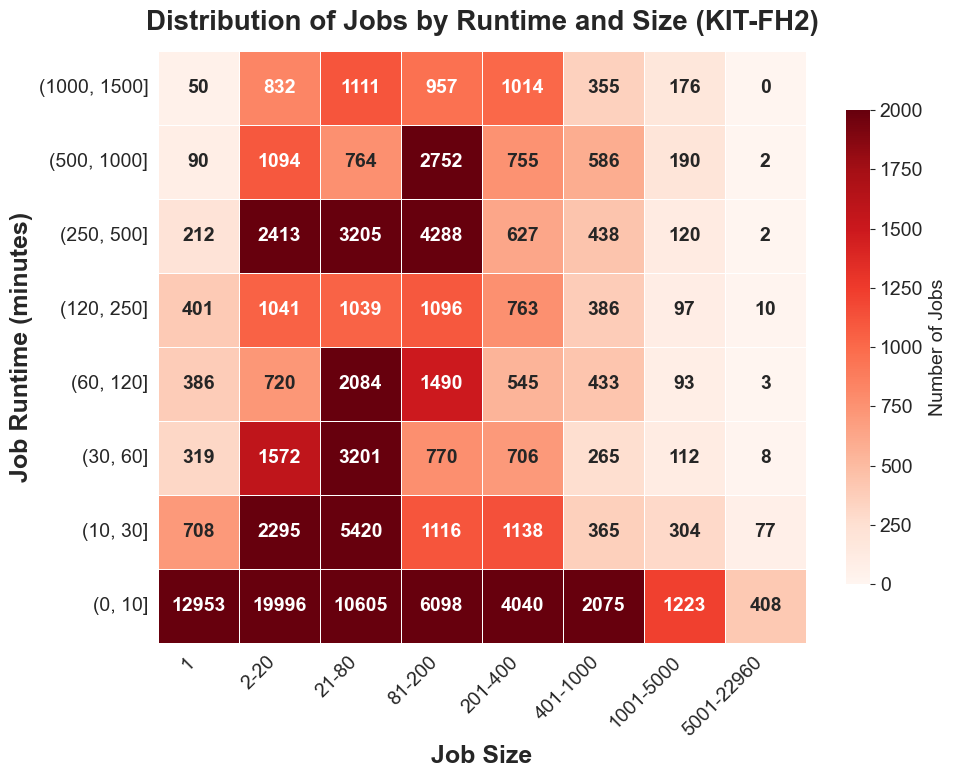}
        \caption{KIT-FH2}
        \label{fig:heat-kit}
    \end{subfigure}\hfill
    \begin{subfigure}[t]{0.49\textwidth}
        \centering
        \includegraphics[width=\linewidth]{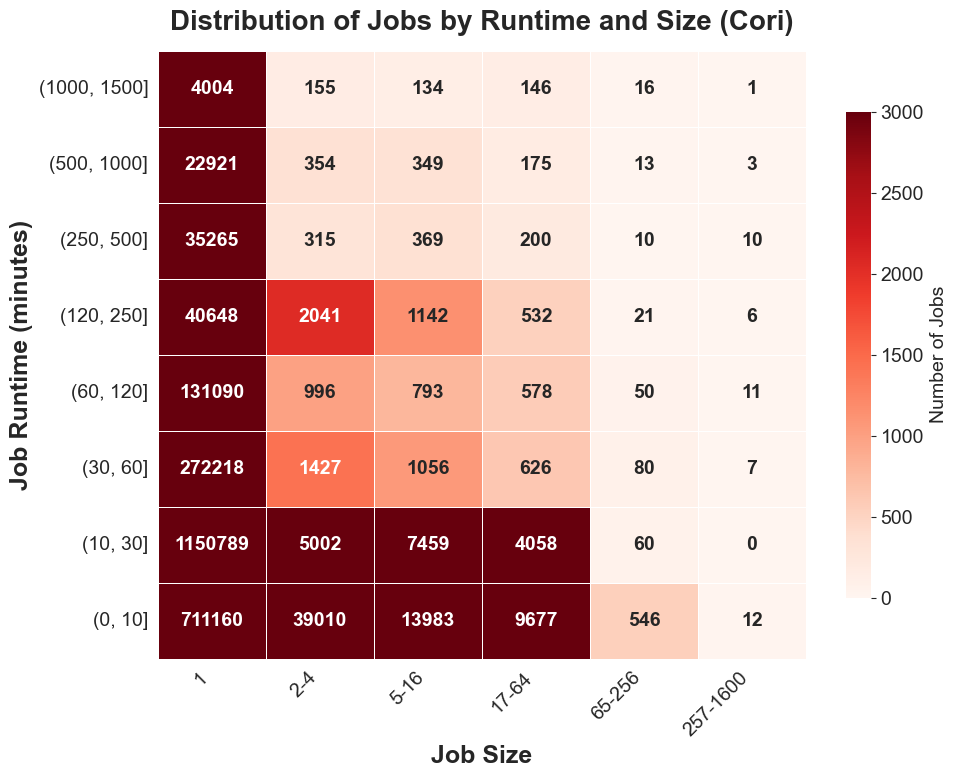}
        \caption{Cori}
        \label{fig:heat-cori}
    \end{subfigure}

    \vspace{2mm}

    \begin{subfigure}[t]{0.49\textwidth}
        \centering
        \includegraphics[width=\linewidth]{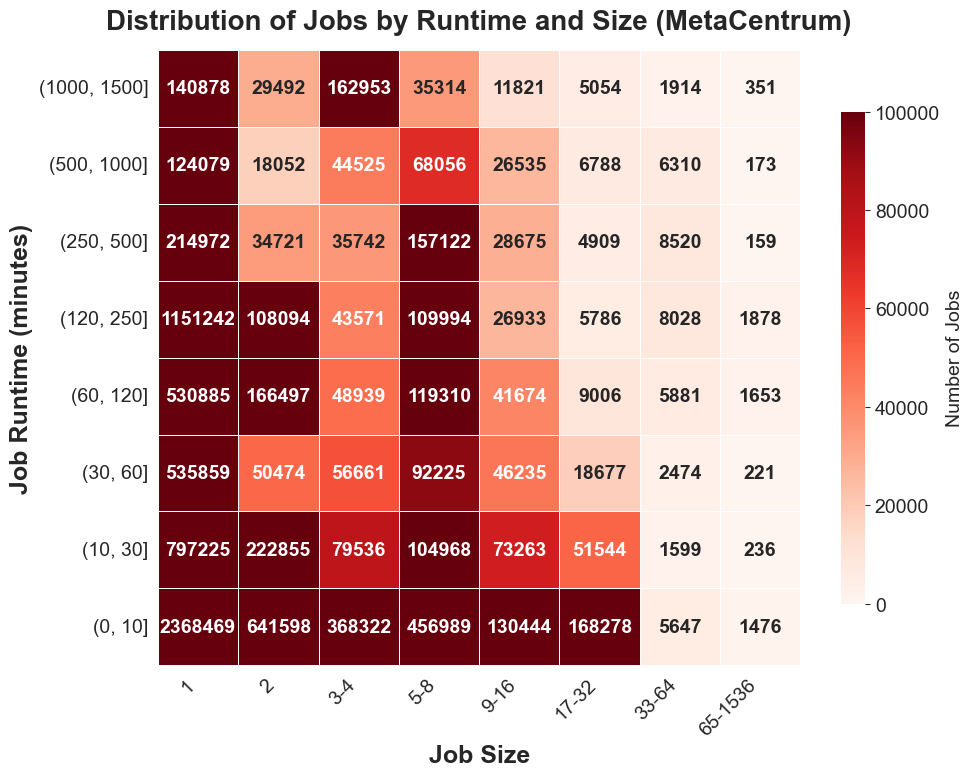}
        \caption{MetaCentrum}
        \label{fig:heat-meta}
    \end{subfigure}\hfill
    \begin{subfigure}[t]{0.49\textwidth}
        \centering
        \includegraphics[width=\linewidth]{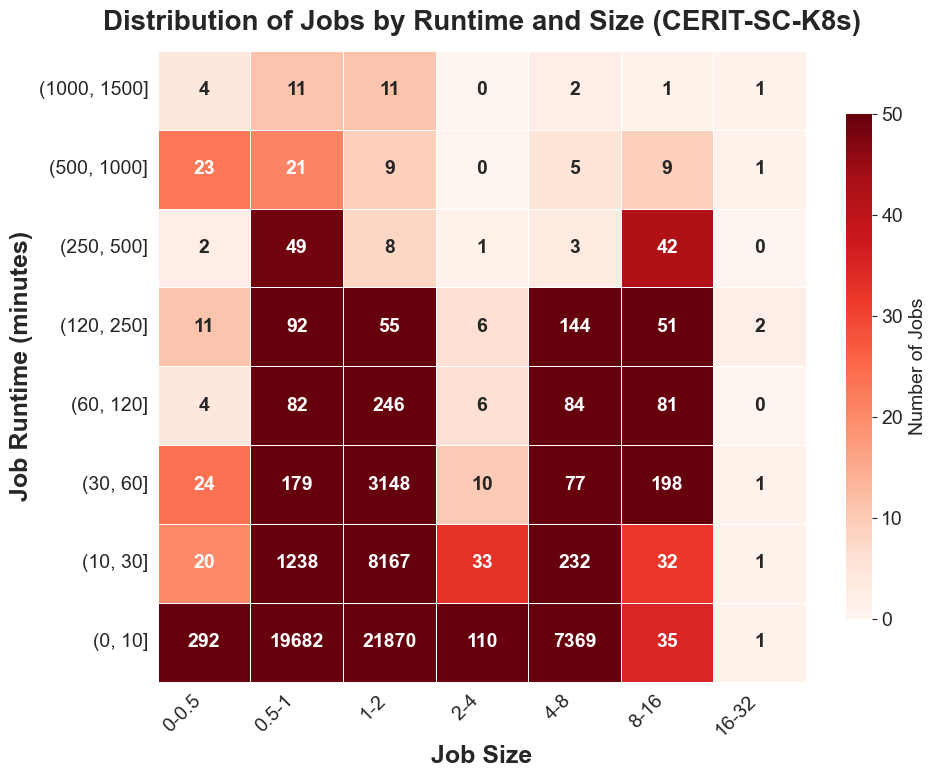}
        \caption{CERIT-SC-K8s}
        \label{fig:heat-cerit}
    \end{subfigure}

    \caption{Joint distribution of job runtime and job size across cross-system traces. Each cell shows the number of completed jobs in a runtime bin (rows) and size bin (columns); color intensity is capped for visual clarity.}
    \label{fig:heatmap-all-traces}
\end{figure}

\subsection{Cross-System Robustness}
\label{subsec:robustness}

To demonstrate that SafeWall generalizes beyond a single production system, we evaluate it on four additional public workload traces spanning a wide range of system scales, architectures, and scheduling paradigms (Table~\ref{tab:other_traces}). All traces are obtained from the Parallel Workloads Archive~\cite{pwa_site}, NERSC~\cite{nerscCoriDocs}, JSSPP Workloads~\cite{klusacek2023softwalltime,klusacek2024trace}, and differ substantially in workload composition and user behavior.

Cori~\cite{nerscCoriDocs} is a Cray~XC40 system at the National Energy Research Scientific Computing Center (NERSC) comprising both Intel Knights~Landing and Haswell partitions. Its 2022 trace contains approximately 2.46 million jobs submitted by 1{,}318 users, representing a large-scale leadership-class HPC workload.

KIT-FH2~\cite{pwa_site} corresponds to the ForHLR~II cluster at the Karlsruhe Institute of Technology (1{,}173 nodes, 24{,}048 cores). Its 19-month trace includes about 114{,}000 jobs from 164 users and exhibits a notable 6.6\% censoring rate, indicating frequent walltime exhaustion.

MetaCentrum~\cite{klusacek2024trace} is the Czech National Grid Infrastructure, a federated environment spanning 47 clusters and 34{,}400 cores across 42 queues. Its 2023 annual log is the largest dataset in our study, with over 10.1 million jobs from 1{,}618 users and a 4.9\% censoring rate.

CERIT-SC-K8s~\cite{klusacek2023softwalltime} is a 20-node Kubernetes cluster at CERIT-SC (Czech Republic), where each workload unit corresponds to a single-node container pod. Its 11-month trace comprises roughly 64{,}000 pods from 103 namespaces and represents a containerized scheduling paradigm distinct from traditional batch systems.

Despite their architectural and operational diversity, all four traces exhibit the heavily right-skewed runtime distributions characteristic of HPC and cloud workloads. Median runtimes are short---approximately 17~minutes for Cori, 10~minutes for KIT-FH2, 22~minutes for MetaCentrum, and roughly 3~minutes for CERIT-SC-K8s---yet mean runtimes are 3--67$\times$ larger than the respective medians, reflecting long tails of resource-intensive jobs that dominate total consumption. Skewness ranges from 4.1 on KIT-FH2 to 31.6 on CERIT-SC-K8s, underscoring the prevalence of extreme runtimes across systems. Figure~\ref{fig:heatmap-all-traces} further visualizes this pattern via the joint runtime--size distributions, where the mass concentrates in short, small jobs with a sparse but important long-duration/large-scale tail.

User walltime behavior also varies substantially across traces. MetaCentrum users are particularly conservative, with a median runtime-to-walltime ratio of only about 3.5\%, compared with roughly 36.5\% observed on Theta; KIT-FH2 falls between these extremes. These differences in utilization patterns, censoring rates, and scheduling models (batch, grid, and Kubernetes) create a challenging generalization setting for any walltime recommendation method.

We apply the same two-component pipeline on all four traces: first training a point runtime predictor using submission-time features, then calibrating a nonnegative SafeWall buffer using training residuals only. Across all datasets, the base predictor achieves strong predictive fit (Table~\ref{tab:robustness_combined}), yet the unbuffered underestimation rate remains close to 50\%, consistent with the behavior of symmetric-loss regression under skewed error distributions. SafeWall consistently reduces underestimation risk across systems while preserving predictive fidelity, demonstrating robustness to differences in scale, workload mix, and scheduling paradigm.

\subsection{Analyzing the Impact of Risk-Aware Buffering}
\label{subsec:user_bias}
Prediction errors are not uniformly distributed across users. Among ``stable'' test users (those with at least 10 jobs), 35.8\% exhibit a positive mean residual (systematic under-prediction) and 64.2\% exhibit a negative mean residual (systematic over-prediction). Moreover, user mean-residual bias shows persistence from train to test (Pearson correlation 0.167 among stable users). This empirical heterogeneity motivates risk-aware buffering: a single global margin is an average-case policy that can remain unsafe for some users while being unnecessarily conservative for others.

\section{Discussion}
\label{sec:discussion}

SafeWall is motivated by a practical mismatch between how runtime predictors are typically trained and how their outputs are used as walltime limits in batch scheduling. Our results suggest that explicitly separating \emph{estimation} (predict $\hat{y}$) from \emph{risk control} (add a nonnegative margin $b$) is an effective way to convert strong base predictions into scheduler-safe recommendations. Below, we summarize the main strengths, limitations, and implications for deployment.

\subsection{Strengths}

SafeWall's main strength is \emph{its modular two-component design}. The buffer adapter relies only on historical residuals and can wrap any base predictor, including future models with richer features, without altering the safety logic. Improvements in prediction quality directly benefit, while the risk-aware buffering layer provides a separate, auditable mechanism for controlling underestimation risk.

From an operational standpoint, SafeWall is \emph{simple and tunable}. A single parameter (e.g., a target underestimation rate or residual percentile) directly controls the trade-off between recommendation tightness and slack. Because buffering is applied post-training, SafeWall remains \emph{model-agnostic}, allowing for straightforward validation and deployment. Moreover, risk-aware buffers offer a practical balance between global and personalized policies: it adapts to individual user behaviors when historical data are abundant, while employing shrinkage toward a global baseline to mitigate instability in data-sparse or cold-start scenarios.

\subsection{Limitations}
SafeWall's calibration depends on the semantics and quality of runtime labels. In real scheduler logs, jobs that reach their requested walltime are right-censored: the recorded runtime reflects the walltime limit rather than the job's true completion time. If such walltime hits are common, then residuals computed from observed runtimes can understate the positive tail (i.e., the most harmful underpredictions), which may lead to buffers that are not conservative enough. In these settings, censored records should be handled explicitly during calibration. Two practical options are: (i) treat walltime-hit jobs as incomplete observations and incorporate them as conservative lower bounds when estimating tail residuals (so the buffer does not rely on artificially capped runtimes), or (ii) leverage refined-runtime or soft-walltime traces that reconstruct more accurate runtimes for censored jobs and thus provide cleaner residual signals for buffer estimation~\cite{klusacek2024trace}.

SafeWall assumes residual behavior is \emph{reasonably stable} between calibration and use.
Significant shifts in workload composition, software environments, or scheduling policies can alter residual distributions and degrade calibration accuracy.
While rolling-window updates can mitigate such drift, they introduce additional hyperparameter tuning requirements.

Finally, personalization may introduce \emph{unequal service levels}, as harder-to-predict users receive larger buffers. Depending on site policy, group-based buffering policies (per partition, project, or job class) may be appropriate to ensure equitable resource access.

\section{Conclusion}
\label{sec:conclusion}

This work addressed the practical gap between \emph{accurate runtime prediction} and \emph{safe walltime recommendation} in batch-scheduled HPC systems.
Using the 6.5-year Theta log, we showed that even strong symmetric predictors can yield unacceptably high underestimation rates under hard walltime limits.
To address this, we introduced SafeWall, a two-component approach that separates estimation from risk control by augmenting a base predictor with a model-agnostic, risk-aware buffering layer that explicitly manages one-sided termination risk.
Experiments on the Theta trace and multiple public datasets demonstrate that SafeWall outperforms existing methods by offering a well-defined accuracy-risk trade-off; specifically, its residual-calibrated buffers provide predictable operating points, while the risk-aware buffering achieves lower underestimation rates than global or prior conservative approaches without introducing unnecessary slack.
Overall, the results support treating walltime recommendation as an explicit risk-control problem.

SafeWall provides a modular framework that complements existing runtime prediction models. As a lightweight solution relying solely on submission-time features and historical data, it can be deployed as either a standalone recommendation service or a submission-time wrapper that automatically adjusts walltime requests. The source code is publicly available at \url{https://github.com/SPEAR-UIC/JSSPP26-SafeWall}.

Future work includes making calibration conditional on job attributes (e.g., queue/partition, node count, or project) so the buffer adapts to systematic heterogeneity beyond users, and integrating SafeWall into a scheduler workflow to evaluate system-level impacts such as backfilling efficiency, utilization, and wait time. We also plan to handle right-censored jobs explicitly in environments where walltime hits are frequent, and to pair SafeWall with uncertainty-aware base predictors (e.g., quantile or conformal calibration) to strengthen reliability while preserving the same modular two-component design.

\bibliographystyle{splncs04}
\bibliography{references}

\end{document}